# In Situ Ion Induced Gelation of Colloidal Dispersion of Laponite: Relating Microscopic Interactions to Macroscopic Behavior

Rashmi P. Mohanty, Khushboo Suman and Yogesh M. Joshi*

Department of Chemical Engineering and Center of Environment Science and Engineering, Indian Institute of Technology Kanpur 208016, India

* Corresponding author, E-Mail: joshi@iitk.ac.in,

Tel: +91 512 259 7993 Fax: +91 512 259 0104

**Abstract**

Aqueous dispersion of Laponite, when exposed to $CO_2$ environment leads to in situ inducement of magnesium and lithium ions, which is, however absent when dispersion is exposed to air. Consequently, in the rheological experiments, Laponite dispersion preserved under $CO_2$ shows more spectacular enhancement in the elastic and viscous moduli as a function of time compared to that exposed to air. By measuring concentration of all the ions present in a dispersion as well as change in pH, the evolving inter-particle interactions among the Laponite particles is estimated. DLVO analysis of a limiting case is performed, wherein two particles approach each other in a parallel fashion – a situation with maximum repulsive interactions. Interestingly it is observed that DLVO analysis explains the qualitative details of an evolution of elastic and viscous moduli remarkably well thereby successfully relating the macroscopic phenomena to the microscopic interactions.

**Keywords:** Clay dispersion; Laponite dispersion; rheology; DLVO theory; inter-particle interactions.

1. Introduction

The macroscopic properties of the colloidal dispersions are determined by various factors that include the nature, the shape and the size distribution of the particles, their concentration, the distribution of charges on the same, etc. [1]. These factors in turn determine the entropic considerations as well as the energetic interactions among the particles. With respect to the energetic interactions, while the change in pH determines the nature and the magnitude of the charges on a particle surface, gradual increase in the ionic concentration leads to progressive shielding of the surface changes [2]. In the present work a colloidal dispersion of Laponite, a disk shaped nanoparticle, in water is studied wherein many of the above mentioned factors are present. An aqueous dispersion of water shows spectacular enhancement in viscosity and elasticity as a function of time, and origin of this behavior has been a subject of intense investigation over past two decades [3-8]. However, unlike many naturally occurring clays, Laponite is vulnerable to chemical degradation when exposed to acidic environment [9-11], usually caused by dissolution of atmospheric carbon dioxide. Under such conditions, Laponite releases various kinds of ions in the aqueous media in which it is suspended [11]. In addition, pH of the dispersion also shows time dependent evolution. Both these in situ effects gradually alter the electrostatic interactions among the Laponite particles.

Laponite is 2:1 is a synthetic clay mineral belonging to a family of hectorite clays with a chemical formula of the unit cell: $Na^{+0.7}[(Si_8Mg_{5.5}Li_{0.3})O_{20}(OH)_4]^{-0.7}$ [12]. In Laponite two sheets of tetrahedral silica sandwich octahedral sheet of magnesia leading to a single crystal of Laponite having disk like shape with 25±2.5 nm diameter and 0.92 nm thickness [13]. In the middle sheet, lithium isomorphically substitutes magnesium leading to deficiency of positive charge rendering outer two faces of Laponite a negative charge, which is compensated by sodium ions. The edge of a Laponite disk is composed of broken crystal and is dominated by Mg–OH, whose point of zero charge is around 12.5 [14]. Consequently, below this pH, the edge acquires a positive charge. As a result, in an aqueous medium with pH around 10, Laponite particles share edge – to – face attraction while face – to – face repulsion. Both these interactions contribute to structural build up in Laponite dispersion, which causes viscosity and elasticity of the same to show spectacular increase when Laponite is dispersed in water [15]. Incorporation of salt in aqueous dispersion of Laponite is observed to enhance the rate at which viscosity and elasticity of the same increases [16]. Such accelerated increase has been attributed to shielding that the dissociated ions provide to the charged faces of a disk, thereby reducing repulsion among the particles. This causes reduction in the repulsive barrier height for the particles to approach each other leading to faster structural build-up [16, 17].

While Laponite has been in use in variety of industrial applications and in academic studies, Laponite is susceptible to chemical degradation when exposed to acidic environment. Many groups have investigated the chemical stability of hectorite clay mineral over the past few decades [18-20], wherein $H^+$ ions have been suggested to react with clay causing $Mg^{2+}$ ions to leach. Typically, $H^+$ ions have been proposed to attack the clay particles at two locations: the faces and the edges [18, 20]. It has been conjectured that $H^+$ ions get adsorbed on the negatively charged faces, which eventually penetrate the tetrahedral silica sheet to react with octahedral magnesia sheet [19]. The attack on the edges, instead, is more direct. Since the former case involves two steps – the adsorption followed by the penetration – before the reaction takes place [19], it is believed to be slower than the latter [18, 20]. In the presence of weak acids, on the other hand, it has been claimed that $H^+$ ions primarily attack the edges [18]. Interestingly Kreit et al. [19] reported that presence of salt stabilizes the clay, which they attribute to shielding of the clay particles by cations against $H^+$ ions attack.

The chemical stability of Laponite was investigated for the first time by Thompson and Butterworth [9] who claimed that degradation occurs when pH of a dispersion decreases below 9. They proposed that Laponite reacts with acid to undergo the following reaction [9]:

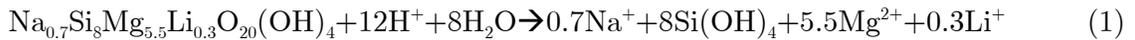
$$Na_{0.7}Si_8Mg_{5.5}Li_{0.3}O_{20}(OH)_4 + 12H^+ + 8H_2O \rightarrow 0.7Na^+ + 8Si(OH)_4 + 5.5Mg^{2+} + 0.3Li^+ \qquad (1)$$

Mourchid and Levitz [10] studied relaxation dynamics of dispersions preserved under air and $N_2$ atmosphere. Interestingly they observed that samples preserved under air underwent gelation while that of kept under $N_2$ atmosphere did not. Mourchid and Levitz [10] attribute this observation to dissolution of $CO_2$ present in air, which causes leaching of $Mg^{2+}$ and $Li^+$ ions enhancing the ionic strength of the dispersions. They proposed that increase in concentration of $Mg^{2+}$ and $Li^+$ ions is responsible for accelerated aging of Laponite dispersion. Jatav and Joshi [11] carried out a systematic study wherein they exposed the Laponite dispersions with varying concentrations of Laponite and salt to air atmosphere. They observed that dispersions with high concentration of Laponite and/or that of salt do not show traces of leached $Mg^{2+}$ ions suggesting that the cations associated with a salt as well as the counter-ions provide a stabilizing effect against $H^+$ ions attack. Very recently Mohanty and Joshi [21] proposed a chemical stability phase diagram which separates the chemically stable region from that of the unstable region as a function of time with respect to concentration of Laponite and that of salt.

The faces of the disk/plate shaped particles of clays experience repulsive interactions due to their negatively charged nature, while experience attraction due to van der Waals interactions. As a result, historically this system has been studied as a model candidate for an application of DLVO theory [22], which accounts for electrostatic as well as van der Waals interactions among the particles [1]. While applying the theory the two negatively charged flat plates are considered to approach each other in a parallel fashion. This parallel configuration, which, while is only one of the possible configurations, provides information about the maximum repulsive interactions, therefore rendering a qualitative insight. In case of Laponite DLVO theory has been employed to infer about how the microstructure of Laponite dispersion gets affected by concentration of Laponite [16, 23], that of externally added salt [16, 23], concentration of counterions [24], temperature [16, 23] externally added polymers leading to steric interactions [25], etc. It has been observed that DLVO theory predicts the qualitative dynamics originating from the interparticle interactions very well.

This work studies effect of $Mg^{2+}$ ion leaching by systematically exposing Laponite suspension to the CO2 environment. Furthermore, spontaneous evolution of Laponite dispersion with and without in situ $Mg^{2+}$ ion leaching is analyzed using rheology and DLVO theory, wherein the latter case involves consideration of multivalent cations. This work, on the one hand studies how the macroscopic rheological behavior changes as a function of time, while on the other hand, how the microscopic inter-particle interactions get altered with time is carefully analyzed. The objective of this work is to relate time evolving microscopic inter-particle interactions to the macroscopic rheological behavior of an aqueous dispersion of Laponite due to in situ change in the ionic concentration.

## 2. Material, sample preparation and experimental procedure

In order to prepare aqueous dispersion, Laponite XLG® (BYK Additives Inc.) was dried for 4 hours at 120°C. Before mixing Laponite with ultrapure water (resistivity 18.2 MΩ·cm) its pH was maintained at 10 by adding NaOH and molarity was fixed at predetermined value by adding NaCl. The mixing of Laponite and water was carried out for 45 min using Ultra Turrex drive, which leads to clear and colorless dispersion. In this work 2.8 mass% Laponite dispersion was studied with no externally added salt (0.1 mM $Na^+$ ion concentration). The freshly prepared dispersions were stored in several polypropylene tubes of (2.74 cm diameter) 50 ml volume so as to have 5 ml of empty space above the dispersion. The samples were purged with $N_2$, air and $CO_2$. 2.8 mass% Laponite dispersion with 5 mM NaCl was also used, and exposed this dispersion to only $CO_2$ environment. The reason for employing this system is discussed in the next section. All the tubes were stored in different desiccators. The free space of the desiccators was also purged with a gas same as that in the free space of tubes. For each experiment a fresh tube was used without disturbing the others. The samples were investigated for a period of 30 days with an interval time of 5 days. This time, over which the sample was kept under quiescent condition, has been represented as rest time $(t_r)$.

Changes in pH and ionic conductivity of a dispersion were measured using Horiba F-71 pH Meter and Horiba DS-71 Conductivity meter respectively. Extent of dissolution of the Laponite particle leading to leaching of $Mg^{2+}$ ions was studied by performing complexometric titration, wherein dispersion is titrated against EDTA (ethylenediamine tetra-acetic acid) using EBT (eriochrome black-T) as the indicator. In the presence of $Mg^{2+}$ ions the dispersion turns red or purple on addition of EBT. To maintain the pH of 10 throughout the titration process, ammonia buffer solution (ammonium chloride/ ammonia) was used. As Laponite forms thixotropic gel, the viscosity/elasticity of the dispersion was first reduced by shearing it vigorously to ease the titration process. The titration's end point occurs when EDTA changes the color of the dispersion to blue, which yields concentration of leached magnesium ion $(C_{Mg})$. The method is sensitive enough to detect $C_{Mg} = 10^{-3}$ mM [26].

The rheological study was carried out using a cup and bob geometry (bob diameter 10.004 mm with 0.407 mm gap) of Anton Paar MCR 501 rheometer. To erase history associated with the samples on a given day after preparation of the same (rest time), pre-shearing was performed by applying large amplitude oscillatory shear of 80000% for 10 minutes. Subsequent to shear-melting, dispersion was allowed to age wherein evolution of elastic modulus $(G')$ and viscous modulus $(G'')$ were monitored by applying small amplitude oscillatory shear of 0.1% for 3 hours. The time elapsed since shear melting has been termed as waiting time or aging time $(t_w)$. Frequency of oscillation was maintained at 0.1 Hz.

Unless otherwise mentioned, measurements of the ionic conductivity, pH, and $C_{Mg}$ reported below are the values averaged over the entire volume of the tube. The rheological study, on the other hand, requires very small amount of sample (around 3 ml), which was collected near the free surface. For 2.8 mass% dispersion studied here, no difference was found in the results when dispersion was stored under the $N_2$ or air atmosphere for all the experiments. Therefore, only the results associated with the air atmosphere are reported below. All the experiments were performed at 25°C.

## 3. Results and discussion

A visual observation of the 2.8 mass% dispersions without any salt preserved under $CO_2$ and air indicates that dispersion stored under $CO_2$ environment undergoes liquid to soft solid transition after a rest time of 5 days. The samples stored under air atmosphere, on the other hand, took over 15 days to show soft solid-like consistency. More quantitatively the structural evolution of Laponite dispersion has been explored using rheology. In Fig. 1, $G'$ and $G''$ are plotted as a function of time elapsed since shear melting $(t_w)$ for experiments carried out on different $t_r$ for 2.8 mass% dispersion with no salt (air as well as $CO_2$ environment) and 5 mM salt ($CO_2$ environment). All the dispersion samples studied in this work were in liquid state immediately after preparation of suspension on $t_r$= 0 day, and the corresponding values of $G'$ and $G''$ were very small compared to the values on $t_r \geq$ 5 day. We have therefore not plotted the $G'$ and $G''$ data for $t_r$= 0 day in Fig 1. For system with no salt preserved under air, Fig. 1(a) indicates that for all the explored $t_r$, that both $G'$ and $G''$ increase with $t_w$. In the limit of small $t_w$, $G''$ dominates over $G'$, but increase in $G'$ is faster than that of $G''$. In due course $G'$ crosses over $G''$; and at high $t_w$, $G''$ starts decreasing after showing a maximum. However, the decrease in $G''$ is not apparent for $t_r$=5 and 10 day as aging in these two systems is slow. Literature suggests that if waited for long enough waiting time, $G''$ of both these systems should also show decrease [16]. This behavior originates from inability of shear melting to completely obliterate evolved structure of aged aqueous dispersion of Laponite due to phenomenon known as irreversible aging [27, 28]. Consequently, at progressively higher $t_r$, unbroken structure remained in a sample after the shear melting causes aging to begin at a mature state shifting the evolution of $G'$ and $G''$ to lower values of $t_w$.

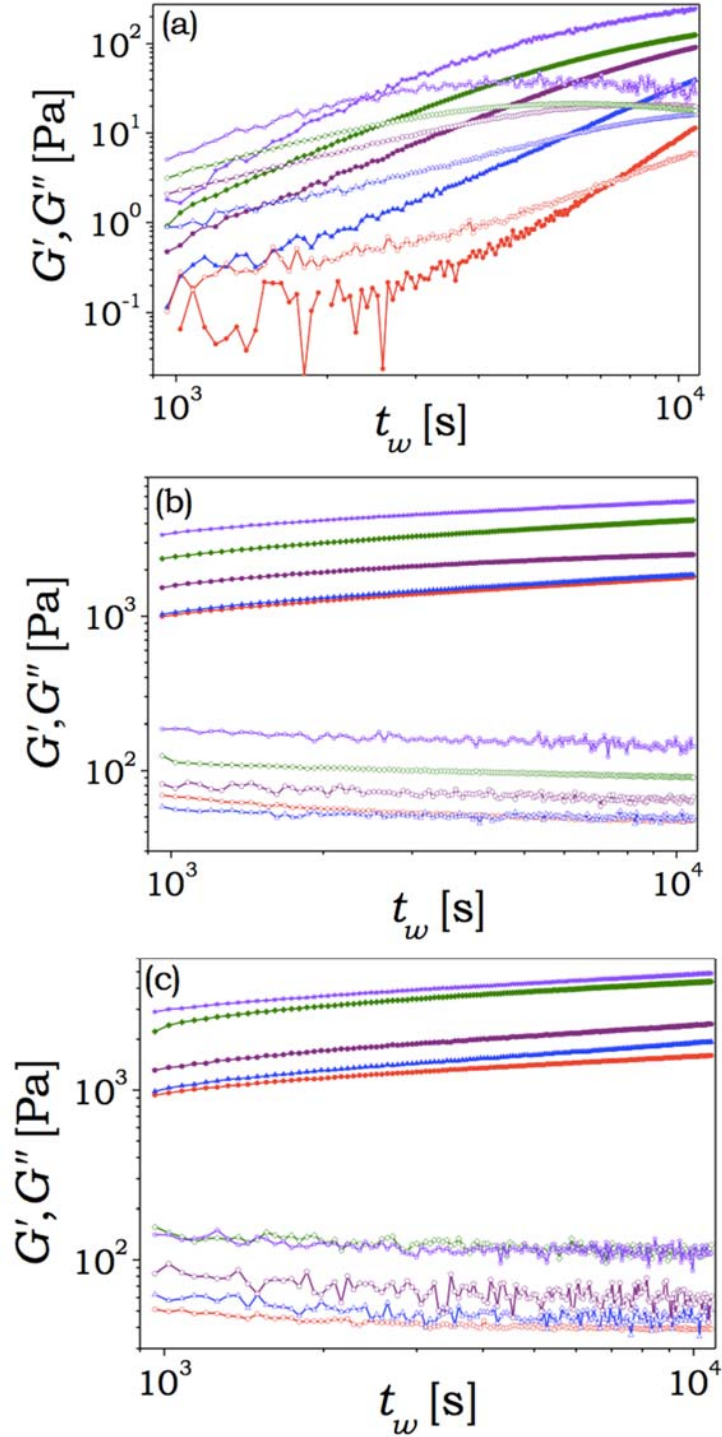

FIG. 1. Elastic modulus ($G'$, by closed symbols) and viscous modulus ($G''$, open symbols) are plotted as a function of time elapsed since shear melting ($t_w$) for 2.8 mass% dispersion with (a) no salt under air atmosphere, (b) no salt under $CO_2$ environment, (c) 5 mM salt under $CO_2$ environment for different rest times (bottom to top): 5 day (circles), 10 day (up triangles), 15 day (pentagons), 20 day (diamonds), 30 day (stars).

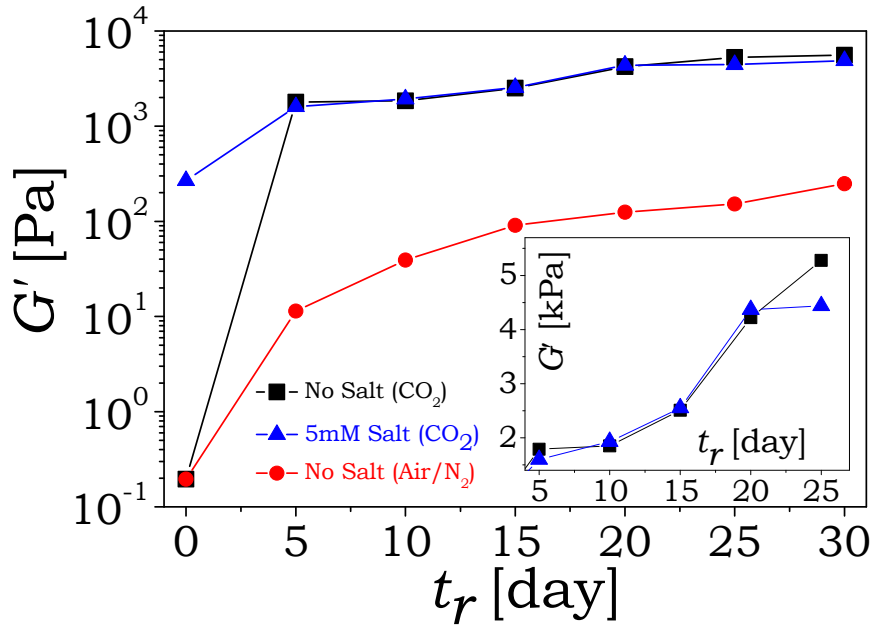

FIG. 2. Storage modulus ($G'$) at $t_w = 3$ h, is plotted as a function of $t_r$ for different Laponite dispersion samples: no salt stored in air (circles), no salt stored in $CO_2$ (squares), 5mM salt stored in $CO_2$ (triangles). In the inset $G'$ is plotted against $t_r$ for the dispersions kept under $CO_2$ for a better clarity. The lines serve as a guide to the eye.

In Fig. 1(b) evolution of $G'$ and $G''$ is plotted for the samples preserved under $CO_2$ environment. However, on $t_r=5$ day itself, in a behavior very different from that shown in Fig. 1(a), system is observed to be in a state with $G'$ over an order of magnitude greater than $G''$. With increase in $t_r$, $G'$ and $G''$ shift to the lower values of $t_w$ (when compared at same values of moduli) showing presence of irreversible aging. In Fig. 1(c) behavior of 2.8 mass% dispersion with 5 mM salt preserved under $CO_2$ environment is plotted, which is qualitatively similar to that shown in Fig. 1(b). In Fig. 2 $G'$ at $t_w=3$ h as a function of $t_r$ is plotted for all the three systems reported in Fig. 1. On $t_r=0$ day, $G'$ of no salt system preserved under air and that of $CO_2$ are identical as expected. On the other hand, the values of $G'$ for the salt containing system is higher than both the no salt systems on $t_r=0$ and $t_w=3$ h. However, over $t_r=0$ to 5 days $G'$ of the no salt system preserved under $CO_2$ increases very spectacularly. It is apparent from Figure 2 that $G'$ of two systems preserved under $CO_2$ are comparable from day 5 to up to 20 days. But, after 20 days, $G'$ associated with the no salt system takes the higher values than $G'$ of 5 mM system as shown in the inset of Fig. 2 (The comparable values of $G'$ for 2.8 mass% 5 mM system preserved under $CO_2$ with that of 2.8 mass% and no salt system preserved under $CO_2$ over the initial period was the reason for employing the former system in this study).

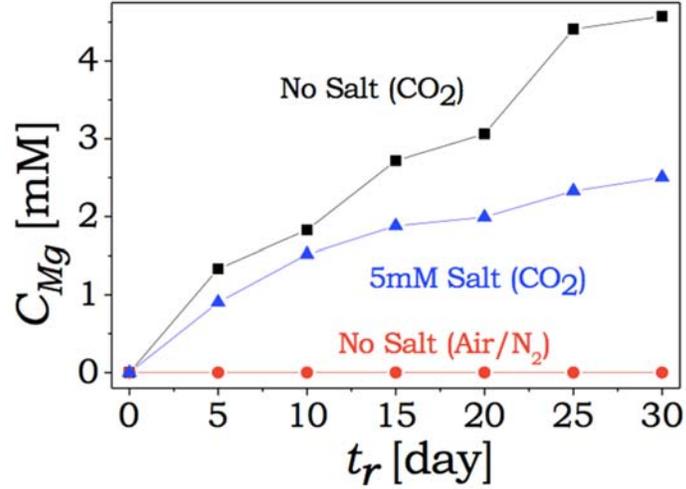

FIG. 3. Average concentration of leached magnesium ions ($C_{Mg}$) in dispersion is plotted as a function of $t_r$ for Laponite dispersion samples: with no salt stored under air (circles), with no salt stored under $CO_2$ environment (squares), with 5mM salt stored in $CO_2$ environment (triangles).

This observation of Fig. 1 and 2 indicates that Laponite dispersion gets affected in $CO_2$ environment causing accelerated structural evolution. Since dissolution of $CO_2$ in Laponite dispersion is known to cause leaching of $Mg^{2+}$ ions from Laponite dispersion, complexometric titration is carried out on the dispersions at various locations away from the free surface. $C_{Mg}$ averaged over the length of tube is plotted as a function of $t_r$ for the studied Laponite dispersion in Fig. 3. The samples preserved under air atmosphere do not show any detectible presence of $Mg^{2+}$ ions. The samples preserved under $CO_2$ atmosphere, conversely, do show significant presence of $Mg^{2+}$ ions. While $C_{Mg}$ increases with increase in $t_r$ in presence of $CO_2$, dispersion having salt shows lesser enhancement in $C_{Mg}$ compared to that having no salt. This observation corroborates with the findings of Joshi and coworkers [11, 21], who claimed that increase in $C_S$ suppresses the creation of acidic environment locally, thereby stabilizing Laponite particles as observed in Fig. 3.

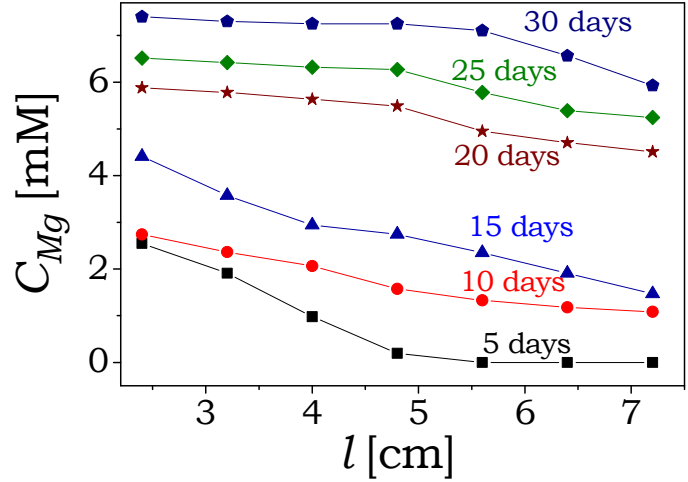

FIG. 4. Concentration of leached magnesium ions ($C_{Mg}$) for a dispersion without salt stored under $CO_2$ environment is plotted as a function of distance from the free surface ($l$) for different $t_r$ (bottom to top): 5 days (squares), 10 days (circles), 15 days (up triangle), 20 days (stars), 25 days (diamonds), 30 days (pentagons).

In Fig. 4, $C_{Mg}$ as a function of distance from the free surface at different $t_r$ is plotted for the dispersion stored under $CO_2$. At small rest times, only the sample closer to the free surface shows dissolution of magnesium. While curve gets shifted to higher values of $C_{Mg}$ with increase in rest time, for the lower $t_r$ (5, 10, and 15 days) the plot indicates decreasing trend of $C_{Mg}$ with the depth of the container. For higher $t_r$ (20, 25, and 30 days), on the other hand, $C_{Mg}$ is observed to vary very weakly near the free surface. The observed behavior is expected as $CO_2$ was purged at the top of the container. The process that leads to leaching of $Mg^{2+}$ ions involves three steps, dissolution of $CO_2$ in aqueous media to form acidic environment, its diffusion to progressive lower depths, and its reaction with Laponite to produce $Mg^{2+}$ ions. In the initial days and deeper depths, the process is diffusion controlled. However, on later days once sufficient dissolution and diffusion has taken place the process is reaction controlled leading to the observed behavior.

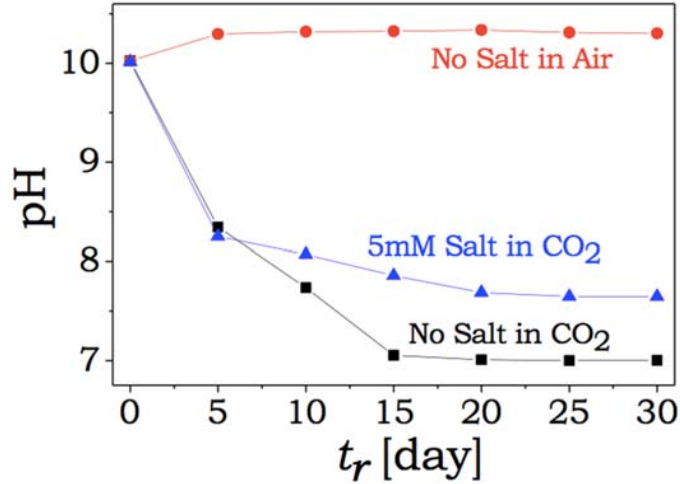

Fig. 5. pH is plotted as a function of $t_r$ for no salt dispersion stored in air (circles), no salt dispersion preserved under $CO_2$ (squares), 5mM salt dispersion preserved under $CO_2$ (triangles).

The results of enhancement in modulus plotted in Fig. 1 and 2 and that of $C_{Mg}$ plotted in Fig. 3 and 4 indicate that greater concentration of $Mg^{2+}$ ions increases the ionic strength of samples preserved under $CO_2$ environment, which leads to accelerated structural evolution. However, prior to causing leaching of $Mg^{2+}$ ions, dissolution of $CO_2$ in aqueous media also leads to decrease in pH of a dispersion, which has multiple effects on the dynamics of structural evolution. On one hand, change in pH affects relative concentration of $H^+$ ions and $OH^-$ ions, on the other hand decrease in pH causes edge of a Laponite particle to become more electropositive enhancing attraction between the edges and the faces. In Fig. 5 pH averaged over the length of a tube is plotted as a function of $t_r$. Immediately after addition of Laponite in water, pH of the same is observed to be 10, the same value that of associated with initial pH of water. However, with increase in $t_r$, pH of sample preserved under air reaches 10.3 within 5 days and remains constant thereafter. The samples kept under $CO_2$ environment, conversely, show decrease in pH as a function of time due to formation of carbonic acid according to [29]:

$$CO_2 + H_2O \leftrightarrow H_2CO_3 \leftrightarrow H^+ + HCO_3^- \qquad (2)$$

The rate of decrease in pH for initial rest time is significant, but it becomes almost constant after a period of 15 days. The decrease in pH is however smaller for a sample containing salt than that for no salt system suggesting presence of salt indeed prohibits formation of acid in a sample.

Leaching of $Mg^{2+}$ ions causes erosion of the Laponite particles. Typically, a single particle of Laponite contains 1000 unit cells. Since each unit cell has around 5.5 Mg atoms, a particle as a whole has 5500 Mg atoms. With knowledge of leached $Mg^{2+}$ ions, as shown in Fig. 3, the extent of erosion of the Laponite particles can be easily calculated. It is suggested by calculations that the largest concentration of leached $Mg^{2+}$ ions per particle for no salt system preserved under $CO_2$ environment on day 30, constitute only 2.16% of the total magnesium ions per particle. As suggested by Kerr et al. [18] it is assumed that for a weakly acidic system, as is the present case (with pH always above 7), edge is more vulnerable for $H^+$ ion attack. Consequently, considering Mg atoms to be evenly placed over the area of a Laponite disk, the calculation suggests erosion of only 2.2% of the area and around 1.1% decrease in the radius of a particle.

The ionic conductivity of sample as a function of rest time is also measured, which is plotted in Fig. 6. For a freshly prepared dispersion without addition of salt, ionic conductivity is observed to be 0.75 mS/cm. For the samples preserved under air, the only contribution to conductivity over and above that associated with $t_r=0$ day is dissociation of $Na^+$ counter ions from faces and change in pH due to protonation of Mg-OH sites on the edges. However, both these contributions induce only marginal increase in conductivity as a function of time as shown in Fig. 6. Ionic conductivity of the dispersions stored under $CO_2$ environment, on the other hand, shows significant increase with rest time. Furthermore, as expected, system having 5 mM salt has higher conductivity at small rest times. However, interestingly, system with no salt preserved under $CO_2$ demonstrates greater increases in conductivity and exceeds that of associated with the former between 10 and 15 days of rest time. Given that increase in conductivity in both these systems is due to leaching of $Mg^{2+}$ ions and change in pH, considering the behavior shown in Fig. 3 and 5 respectively, the trend observed in Fig. 6 is not surprising.

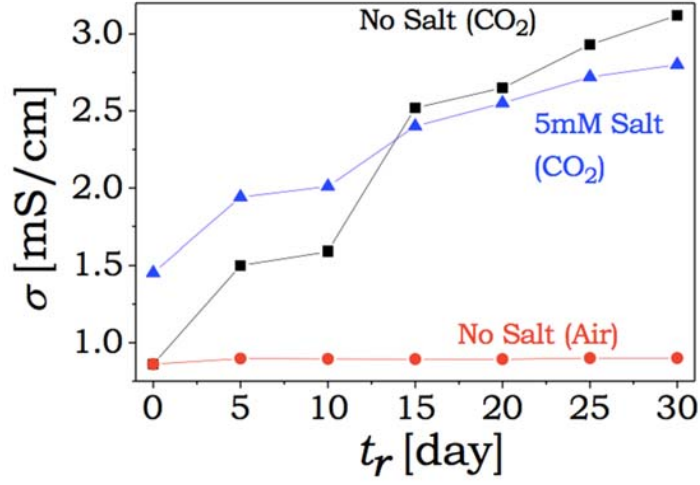

FIG. 6. Ionic conductivity is plotted as a function of $t_r$ for no salt dispersion stored in air (circles), no salt dispersion preserved under $CO_2$ (squares), 5mM salt dispersion preserved under $CO_2$ (triangles).

The ionic conductivity of dispersion depends on concentration and mobilities of ions present in a dispersion given by [30]:

$$\sigma = e\sum_i \mu_i n_i \qquad (3)$$

Where $\sigma$ is the ionic conductivity of dispersion, $e$ is the electron charge, while $\mu_i$ and $n_i$ are respectively the ionic mobility and the concentration (ions/m³) of an $i$ th type of ion. In Laponite dispersion studied in the present work, the various types of ions present are: $H^+$, $OH^-$, $Na^+$, $Mg^{2+}$, $Li^+$ and $Cl^-$. The concentrations of $H^+$ and $OH^-$ ions are calculated respectively from the measured values of pH as: $n_H = 10^{-pH}$ and $n_{OH} = 10^{pH-14}$. The concentration of $Mg^{2+}$ ions ($n_{Mg}$) is obtained from complexometric titration and is mentioned in Fig. 3. The stoichiometry given in equation (1) suggests that for every 5.5 moles of $Mg^{2+}$ leaching, 0.3 moles of $Li^+$ ions are also released. Consequently, estimation of $n_{Mg}$ leads to estimation of $n_{Li}$ as well. The concentration of $Cl^-$ ions is same as amount of NaCl (5mM) added in the dispersion. Mobility of the above mentioned ions is given by Haynes [31]: $\mu_H = 3.623 \times 10^{-7}$ m²/sV, $\mu_{OH} = 2.05 \times 10^{-7}$ m²/sV, $\mu_{Na} = 5.19 \times 10^{-8}$ m²/sV, $\mu_{Mg} = 1.098 \times 10^{-7}$ m²/sV, $\mu_{Li} = 4 \times 10^{-8}$ m²/sV and $\mu_{Cl} = 7.908 \times 10^{-8}$ m²/sV. The only unknown is concentration of $Na^+$ ions $(n_{Na})$, which can be obtained through equation (3) from the measurement of ionic conductivity and with knowledge of the concentration of all the other ionic species. In principle the values of $\mu_i$ mentioned above are the bulk values, and may decrease because of the presence of charged Laponite particles in the media. However, since obtaining the actual values of $\mu_i$ is practically impossible, the bulk values for an estimation of $n_{Na}$ is considered, which gives a lower bound on the same.

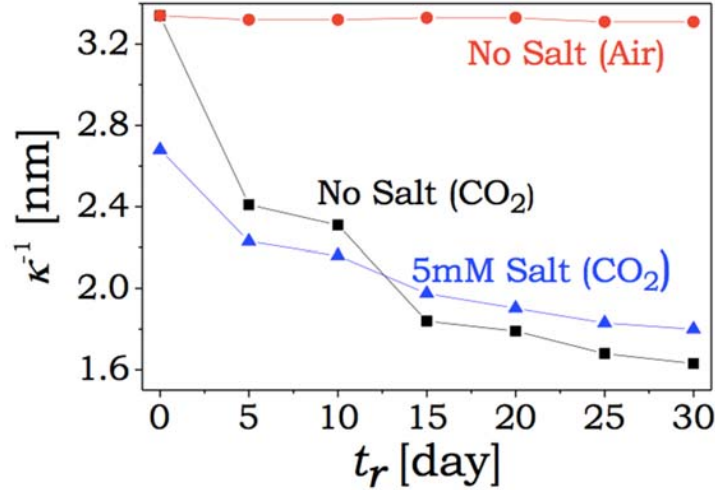

FIG. 7. Debye screening length is plotted as a function of $t_r$ for no salt dispersion stored in air (circles), no salt dispersion preserved under $CO_2$ (squares), 5mM salt dispersion preserved under $CO_2$ (triangles).

With knowledge of the concentration of the each ion present in the dispersion, Debye screening length ($\kappa^{-1}$) of the electric double layer (EDL) associated with the faces of Laponite particles can be calculated by [1]:

$$\kappa^{-1} = \sqrt{\frac{\varepsilon k_B T}{e^2 \sum_i v_i^2 n_i}}, \qquad (4)$$

where, $v_i$ is the valence of the $i$th ion, $k_B$ is Boltzmann constant, $T$ is absolute temperature and $\varepsilon$ is static permittivity of water. The variation of Debye screening length is plotted in Fig. 7. For the dispersion without any salt stored under air, $\kappa^{-1}$ almost remains constant. For the other two systems (preserved under $CO_2$ environment), on the other hand, $\kappa^{-1}$ shows a remarkable decrease with $t_r$. This decrease can be attributed to greater concentration of leached $Mg^{2+}$ ions and decrease in pH. For a system with 5 mM NaCl stored under $CO_2$ atmosphere, owing to the presence of higher amount of $Na^+$ and $Cl^-$ ions in the dispersion, the initial value of $\kappa^{-1}$ was found to be less than that for dispersion prepared without any salt (irrespective of the storage condition). However, with increase in rest time, decrease in $\kappa^{-1}$ is observed to be less intense compared to that for a system with no salt preserved under $CO_2$ atmosphere due to greater leaching of $Mg^{2+}$ ions in the latter.

It is important to note that Debye screening length is not the only factor associated with the faces of Laponite particles, which determines their interactions with the neighboring particles. The dissociation of Na$^+$ ions from the faces of the particles, in addition to the nature and concentration of ions present in the EDL, influence the electrostatic surface potential of the same. The combined effect of the surface potential and the Debye screening length is given by the DLVO theory (Derjaguin-Landau-Verwey-Overbeek) theory [1]. In order to solve the DLVO theory, two plates to approach each other in a parallel fashion is considered and free energy of interaction between them is evaluated as a function of half distance $d$ between the same. The DLVO theory expresses the total free energy per unit area as a sum of that due to the van der Waals interaction $(W_{vdW})$ and the double-layer repulsion $(W_{DL})$.

$$W = W_{vdW} + W_{DL} \tag{5}$$

The free energy per unit area due to van der Waals interaction is given by [22]:

$$W_{vdW}(d) = -\frac{A_H}{48\pi}\left[\frac{1}{(d)^2} + \frac{1}{(d+\Delta)^2} - \frac{8}{(2d+\Delta)^2}\right], \tag{6}$$

where $A_H$ is the Hamaker constant and $\Delta$ is the thickness of unit sheets measured between the same planes (for Laponite $A_H = 1.06 \times 10^{-20}$ J and $\Delta = 6.6$ Å).

The free energy per unit area between the two plates due to electrical double-layer repulsion is straightforward to obtain when dielectric media contains ionic species of only single valency. However, the present system contains different kinds of ionic species with monovalent and divalent valencies for which nonlinear Poisson-Boltzmann equation does not yield the analytical solution. For such case Israelachvili [1] suggested that free energy per unit area due to electrical double-layer repulsion can be approximated by an expression for a system containing monovalent ions (Na$^+$ ions) that employs the correct Debye screening length obtained by considering effect of all the ions [given by equation (4)]. The approximated expression for $W_{DL}$ is therefore given by [1]:

$$W_{DL} \approx \frac{64 n_{Na} k_B T}{\kappa} \gamma^2 e^{-2\kappa d} = W_{DL,\max} e^{-2\kappa d}, \tag{7}$$

where, $n_{Na}$ is the total concentration of Na$^+$ ions in a dispersion obtained by solving equation (3), and $\gamma = \tanh(\Phi_0/2)$. The dimensionless surface electric potential, $\Phi_0$ is related to surface charge density $\varsigma$ by:

$$\varsigma \approx \sqrt{8\epsilon_0 \epsilon_r k_B T n_{Na}} \sinh(\Phi_0) \tag{8}$$

The approximated forms of both the equations (7) and (8) are further justified by the fact that $n_{Na}$ is around an order of magnitude greater than $n_{Mg}$ as shown in Fig. 8.

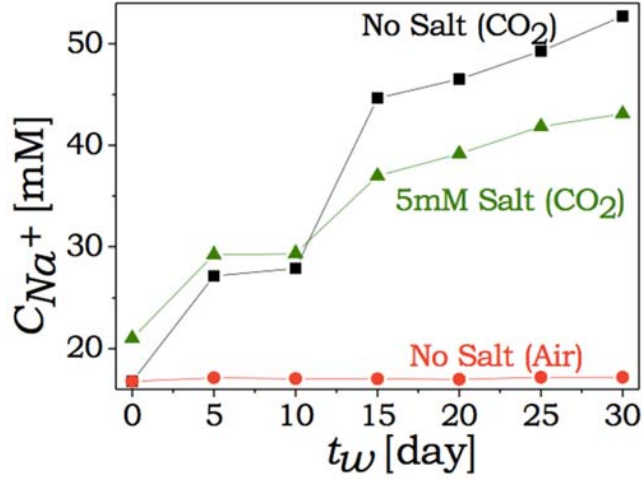

FIG. 8. Average concentration of $Na^+$ ions ($C_{Na}$) in suspension is plotted as a function of $t_r$ for Laponite suspension samples: with no salt stored under air (circles), with no salt stored under $CO_2$ environment (squares), with 5mM salt stored in $CO_2$ environment (triangles).

The charge on the Laponite surface is solely due to dissociation of $Na^+$ ions from the same. Consequently, knowing concentration of dissociation of $Na^+$ ions per face of Laponite particle, $\varsigma$ can be obtained as:

$$\varsigma = e\left(\frac{n_{Na} - n'_{Na}}{2 A_L m_p}\right), \quad (9)$$

where $n'_{Na}$ is the concentration of $Na^+$ ions originating from the sources other than Laponite (NaCl and NaOH), $A_L$ is the area of a face of Laponite, factor 2 represents two faces per particle and $m_p$ are the number of Laponite particles per unit volume. In equation (9), the effect of adsorption of any cations on the negatively charged faces is neglected. $W_{DL,\max}$ and $W_{DL}(d)$ are obtained by solving equations (7) to (9), while by solving equation (6) for $W_{vdW}(d)$, $W(d)$ is obtained.

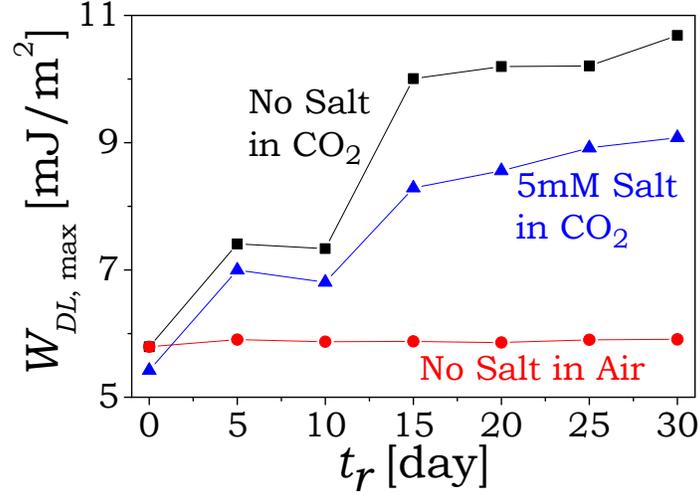

FIG. 9. The leading term of equation (7), $W_{DL,\,max}$ is plotted as a function of $t_r$ for no salt dispersion stored under air (circles), no salt dispersion preserved under $CO_2$ (squares), 5mM salt dispersion preserved under $CO_2$ (triangles).

In Fig. 9 $W_{DL,\,max}$, the leading order term of equation (7) is plotted, that describes the maximum height of the repulsive energy barrier due to only double layer interactions, as a function of $t_r$ for the three studied systems of Laponite dispersion. $W_{DL,\,max}$ associated with a dispersion with no salt preserved under air remains practically constant over an entire duration. On the other hand, for a dispersion with no salt but preserved under $CO_2$, $W_{DL,\,max}$ increases with $t_r$. For a dispersion with 5 mM salt preserved under $CO_2$, $W_{DL,\,max}$ increases with $t_r$ but the increase is not as significant as that of no salt system preserved under $CO_2$.

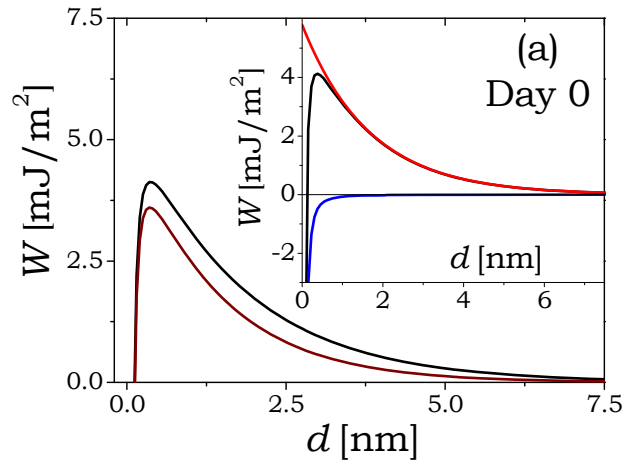

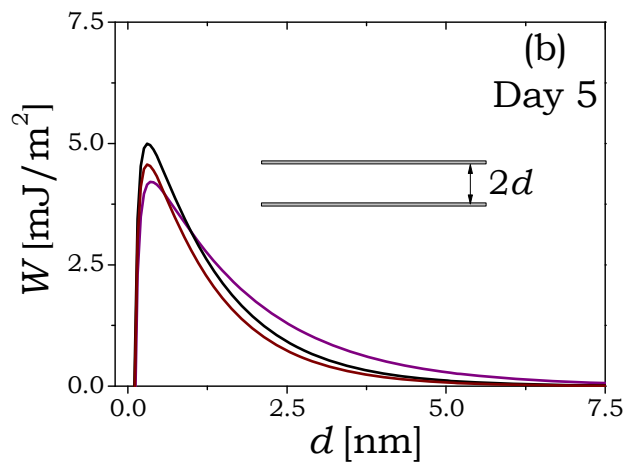

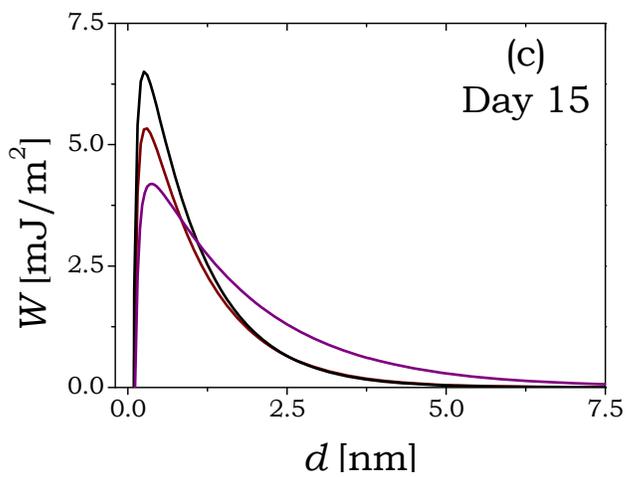

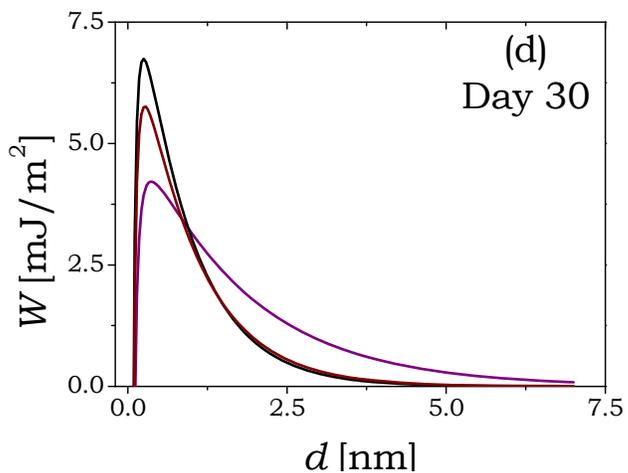

FIG. 10. Variation of total free energy of interaction plotted against the half-distance $d$ between the Laponite plates (refer to the inset of b) for $t_R = 0$ day (a), for $t_r = 5$ day (b), 15 day (c), and 30 day (d). For Fig. (a) the top curve represents dispersion with no salt stored in air as well as $CO_2$, while the bottom curve represents

dispersion with 5mM salt preserved under $CO_2$. For remaining Fig., respectively from the smallest peak to the largest peak: dispersion with no salt stored in air, dispersion with 5mM salt preserved under $CO_2$, dispersion with no salt stored under $CO_2$. The Inset in (a) indicates free energy contribution from van der Waals interactions, double layer repulsion, and their addition as a function of $d$ for no salt dispersion on 0 day.

The complete double layer repulsion term $W_{DL}$ along with $W_{vdW}$ and $W$ as a function of $d$ is plotted in the inset of Fig. 10(a). Both the double layer repulsion ($W_{DL}$) as well as van der Waals attraction $(-W_{vdW})$ increases with decrease in $d$. The sum of both the contributions first increases followed by a sharp decrease with decrease in $d$. While van der Waals attraction prevails over an interparticle distance below 1 nm, double layer repulsion can be seen to be affecting over 5 nm. In addition to van der Waals and double layer interactions, interactions due to ion – ion correlations also play an important role particularly below the interparticle distance of 1 nm [32, 33]. However, since there is no universally accepted free energy expression for the same, its effect has not been explicitly incorporated in equation (5). Furthermore, usually the interactions due to ion-ion correlations are repulsive when monovalent ions are present, and attractive when divalent ions are present in between the plates [33]. Since the present system contains both the types of ions, the cumulative effect is expected to be combination of both attraction as well as repulsion. Interestingly, the van der Waals interactions also show effect over the same length-scale and the double layer repulsion persists between the faces over 5 nm (For the studied Laponite concentration of 2.8 mass %, the average interparticle distance is 35 nm). Therefore, in a parallel configuration the particles experience double layer repulsion over much longer length-scale before they experience the ion-ion correlation. As a result, it is expected that the qualitative effect of ion-ion correlations will not alter the interfacial interactions very significantly. Nonetheless with increase in $Mg^{2+}$ ion concentration, the attractive interactions between the plates in a parallel configuration will certainly augment.

In Fig. 10(a) to (d) $W$ as a function of $d$ is plotted for the three studied systems for $t_r=$ 0, 5, 15, and 30 day respectively. On $t_r=0$ day, the behavior of dispersion without salt preserved under air and $CO_2$ atmosphere is identical. Furthermore, for $t_r=0$ day the height as well as the width of the repulsive barrier between approaching surfaces for dispersion with 5 mM salt stored under $CO_2$ is smaller compared to the no salt system as corroborated by figures 7 and 9. With increase in $t_R$ both the height as well as the width of repulsive barrier between approaching surfaces for dispersion without salt stored under air changes marginally compared to no salt dispersion stored under $CO_2$. The latter shows significant increase in barrier height while decrease in its width. The dispersion with 5 mM salt stored under $CO_2$ also shows increase in barrier height and decrease in its width with increase in $t_R$, but both the changes are not as intense as that of for no salt dispersion stored under $CO_2$.

The behavior shown by Fig. 10 is applicable when two Laponite particles approach each other in a parallel fashion as shown in the inset of Fig. 10(b). However, in reality various Laponite particles can approach each other in any orientation and parallel fashion is just one of the possibilities. Consequently, for a no salt system preserved under air, while undergoing Brownian motion, when two particles approach each other in a non-parallel fashion, at progressively higher $t_r$, attraction between more electronegative face and electropositive edge with decreased width of repulsive barrier accelerates formation of the structure via edge - to - face attractive bonds. The accelerated structure formation is responsible for shifting of $G'$ evolution to lower $t_w$ as shown in Fig. 1(a). However, since for no salt dispersion stored under air, change in all the three properties, namely the barrier height, its width, and pH is marginal, shift in $G'$ evolution is not spectacular. For a system with no salt preserved under $CO_2$ environment, owing to leaching of $Mg^{2+}$ ions, both the height and the width of the energy barrier respectively becomes larger and narrower with increase in $t_r$ with significantly greater intensity compared to that of preserved under air. Moreover, due to greater concentration of $Mg^{2+}$ ions, attraction due to ion-ion correlations also prevails. On the other hand, increase in $t_r$ causes decrease in pH when preserved under $CO_2$ environment which makes the edges of the particles more electropositive. It is therefore not surprising that all these three effects lead to more spectacular buildup of structure manifested in fast evolution of $G'$ in the no salt system preserved under $CO_2$ environment compared to that of under air as shown in figure 1(b).

Comparison of system with no salt and 5 mM salt preserved under $CO_2$ environment is more involved. On one hand, it is known that presence of salt decreases the width of energy barrier. On the other hand, salt renders a stabilizing effect to a dispersion so that decrease in pH as well as amount of leached $Mg^{2+}$ ions is less than that of with no salt system preserved under $CO_2$ as shown in Fig. 3 and 5. Figures 6 and 7 show that up to day 15, dispersion with 5 mM salt has greater ionic conductivity than that of with system with no salt, and therefore the faces of the particles in the former possess smaller Debye screening length. Furthermore, Fig. 9 suggests that $W_{DL,\max}$ is always larger for the dispersion exposed to $CO_2$ environment containing no salt than that of containing 5 mM salt. Smaller Debye screening length originating from greater ionic conductivity suggests reduction in activation barrier to form inter-particle bonds. Due to this effect aggregation rate in salt containing system should be faster than no salt containing system up to day 15. In addition, as concentration of $Mg^{2+}$ ions in the salt free system starts increasing compared to that of salt containing system, the overall attractive interactions among the particles in former due to ion-ion correlation also start influencing the dynamics. Furthermore, the lower pH represents more electropositive edges, while for higher $W_{DL,\max}$, which represents electronegativity of the faces, the rate of aggregation in salt containing system should be lower than no salt containing system throughout the explored rest times. Consequently, due to the competing effects, $G'$ of no salt system shows significant increase initially. After $t_r = 5$ day the values of $G'$ for both the system remain comparable up to 20 days as shown in the inset of Fig. 2. However, beyond 20 days, observations of Fig. 5 to 9, indicate all the factors favor faster structure formation in dispersion exposed to $CO_2$ environment containing no salt than that of containing 5 mM salt. Interestingly, as shown in the inset of Fig. 2, $G'$ of the no salt system indeed takes over $G'$ of dispersion containing 5 mM salt. This observation, therefore, demonstrates a close qualitative correspondence between microscopic electrochemical and macroscopic rheological behavior of aqueous Laponite dispersion in presence of in situ inducement of ions.

In addition to academic aspect, the present work also has important implications from applications point of view. As mentioned before, Laponite is used as a rheology modifier or thickener in variety of industrial products such as home and personal care, mining, agrochemicals, pharmaceuticals, paints, paper, etc. It is highly likely that under usual application conditions, the aqueous products containing Laponite may get exposed acidic environment. Under such conditions, on one hand the chemical degradation induced by the same will lead to erosion of the Laponite particles. On the other hand, it releases various kinds of ions in the aqueous media in which Laponite is suspended altering electrostatic interactions among the particles. Both these effects are not desirable for an intended end use of the product containing Laponite particles. The present work demonstrates such possibility under controlled conditions.

## 4. Conclusion

In this work the interrelation between macroscopic rheological behavior and microscopic electrochemical properties are studied for an aqueous dispersion of Laponite, which demonstrates in situ change in ionic concentration when preserved under $CO_2$ atmosphere. In general, in the rheological experiments carried out on a latter day $(t_r)$ after preparation, evolution of $G'$ subsequent to shear melting shows accelerated evolution. The acceleration is observed to be very spectacular for dispersion preserved under $CO_2$ environment. The complexometric titration and conductivity measurements reveal significant in situ leaching of $Mg^{2+}$ ions and high level ionic conductivity in dispersions preserved under $CO_2$ environment. On the other hand, in comparison, a dispersion preserved under air shows modest ionic conductivity and no traces of leached $Mg^{2+}$ ions. Furthermore, the pH of dispersion preserved under $CO_2$ decreases substantially unlike the pH of preserved under air that shows only slight increase. Knowledge of ionic conductivity and concentration of various participating ions leads to the surface charge density and Debye screening length associated with the faces of Laponite particle in dispersions preserved under various conditions. Also the DLVO theory for two Laponite particles approaching each other in parallel fashion is studied. Very interestingly, analysis of the DLVO results, in addition to the effect of pH on the edge charge of Laponite, explain the rheological behavior of the dispersion very well relating macroscopic behavior to microscopic inter-particle interactions.

**Acknowledgment**

The financial support from the department of Atomic Energy- Science Research Council, Government of India (2012/21/02-BRNS/1097) is greatly acknowledged.